# Right here, right now? The role of spatio-temporal minimum renewable shares for energy system transformation pathways

Tim Fürmann, Ramiz Qussous, Michelle Antretter, Clara Pineau, Rebecca Lovato Franco, Mirko Schäfer
INATECH
University of Freiburg
Freiburg, Germany

*Abstract*—Energy system optimization models are important tools to provide insights regarding trade-offs and interrelations in cost-efficient transformation pathways towards a climate neutral energy system. Using an optimization model of the European electricity system, we study the influence of either an emission cap or temporally resolved minimum renewable generation shares as means to power system decarbonization. Such minimum shares serve as a stylized representation of novel, more granular procurement or certification schemes for renewable power generation. We observe that decarbonization through minimum shares results in a significant increase in system costs compared to the least cost solution obtained through a direct implementation of the corresponding emission reduction as a cap. Nevertheless, these shares provide direct incentives for expanding renewable generation and storage technologies, thus acting as an additional, potentially more robust decarbonization mechanism.

*Index Terms*—Electricity markets, Environmental factors, Low-carbon economy, Renewable energy sources, System modeling

## I. Introduction

The deployment of renewable energy generation technologies is driven by declining technological costs and increasing emission prices, but also by additional support mechanisms like feed-in tariffs or an increasing demand for renewable energy certificates [1]-[3]. In energy system optimization models, the boundary conditions set by the representation of such mechanisms have a significant impact on the details of the resulting scenarios and cost-efficient transformation pathways [4]-[6].

In this contribution, we study the influence of spatio-temporally resolved minimum renewable shares on the cost-efficient system design of a European electricity system under different emission reduction targets. Such minimum shares are a stylized representation of granular "guarantees of origin" or feed-in tariffs, certifying the origin or remunerating the generation of some unit of electricity from renewable sources. The granular nature of this concept includes the condition that this unit must be generated in a certain temporal period and spatial distance with respect to the time and location of consumption, for instance in the same hour and inside the same country. In this context, new schemes like *24/7 clean energy* make use of improved granular greenhouse gas accounting data and suggest time-matched certificates or power purchase agreements [7]-[9]. Only recently, first studies assessing the impact of such granular schemes on the European and US energy system have been published, with a wider discussion in the scientific literature still missing [10], [11].

The article is structured as follows. In the Methodology section, we briefly describe the power system optimization model and introduce temporally resolved minimum renewable shares as boundary conditions. The subsequent section reviews the data underlying the model representation. In the Results section, findings for different minimum shares and emission reduction targets are discussed. The final section provides a brief conclusion and outlook.

### A. Energy system optimization model

We consider a brownfield power system capacity expansion model implemented in the PyPSA framework [12]. The optimization minimizes total annualized costs through expansion of generation, storage and transmission capacities and cost-optimal operation of the system. Already existing generation and storage capacities thus incur operational costs (in particular, fuel costs), whereas the installation and operation of new capacities additionally involve annualized capacity costs. The optimization runs in parallel over all hours of a full year, corresponding to a system with perfect foresight and thus abstracting from an explicit representation of transformation pathways (through myopic optimization, for instance [13], [14]). The system is optimized under various constraints: Generators can only be dispatched in the limit of their nominal power, with additional constraints on the availability of power generation from wind, solar and hydro due to hourly weather data or inflow patterns. Ramping or unit commitment constraints are not implemented. For storage units, constraints on the nominal power and maximum state of charge must be fulfilled, and the storage state of charge must be consistent with the hourly dispatch or uptake. The nodal power balance at all times enforces energy conservation between an exogenous load at each node, and generation, storage dispatch or uptake, and power flows between nodes. These power flows are either modelled as linear optimal power flows (AC connections) or as controllable flows (DC connections) and are limited by the capacity of the lines, respectively [12].

B. *Emission cap vs. minimum renewable shares*

Without further constraints, the system will only install a limited amount of additional generation since the capacity costs involved outweigh the operational costs for already existing conventional generation capacities. Accordingly, the investment into renewable low carbon generation must be induced by additional constraints. A global emission cap limits the total amount of $CO_2$ associated with the operation of conventional power plants, thus limiting the generation from these sources. To still meet the demand, the system thus expands low carbon generation capacities. Alternatively, a minimum share of renewable generation can be imposed as a boundary condition onto the system, directly forcing the system to expand the corresponding (renewable) generation capacities. It can be shown that in the optimization problem, the dual variables to these approaches can be interpreted as corresponding $CO_2$ prices or as renewable feed-in tariffs, respectively. Whereas a $CO_2$ price is associated with an increase in the operational costs of conventional generation, feed-in tariffs correspond to a subsidy facilitating investment in renewable generation [6], [15].

Generally, minimum renewable shares are implemented as constraints regarding the aggregated generation over the entire time span under consideration and over all nodes in the system:

$$\sum_{s \in S,n,t} g_{n,s,t} \geq \gamma \cdot \left[\sum_{s,n,t} g_{n,s,t}\right]. \quad (1)$$

Here, $g_{n,s,t}$ denotes the generation of type $s$ at time $t$ at node $n$. The set $S$ identifies a specific set of generation types, which by the boundary condition in (1) is forced to provide at least a share $\gamma$ of all aggregated generation. In the following, we focus on the temporal perspective and do not consider spatially resolved minimum shares, but the approach can be generalized to allow these cases [11]. Temporally resolved minimum renewable shares are implemented by partitioning the entire time span into periods $\tau$ (days or individual hours, for instance) and apply the constraint in (1) to each of these periods:

$$\sum_{s \in S,n,t \in \tau} g_{n,s,t} \geq \gamma \cdot \left[\sum_{s,n,t \in \tau} g_{n,s,t}\right] \text{ for all } \tau. \quad (2)$$

For $\gamma = 0.5$ and renewable shares with a temporal granularity $\tau$ of a day, for instance, this constraint implies that over each day at least 50% of generation has to be provided by generation of type $s \in S$.

We assume that storage technologies allow to shift generation of a specific type over time. This is implemented by partitioning the storage uptake, dispatch, and state of charge into contributions from the different generation technologies and assuring energy conservation of each type separately at each time. The optimization process then has the possibility to charge storage capacities with a particular type of generation at a certain moment in time, and then dispatch this type in a later period $\tau$ to support fulfillment of the boundary condition (2).

## II. DATA

The model is based on PyPSA-Eur, an open model dataset of the European electricity system at the transmission network level [16], [17]. The network has been clustered to 37 nodes, representing countries of the ENTSO-E area with one node per country (see Fig. 1). The dataset has been extended to include conventional power plants, storage facilities, and the distribution of renewable generation capacity as in 2021. The hourly electricity demand is based on historical load time series published by ENTSO-E. Nuclear power generation and generation from onshore wind, offshore wind, solar, run of river and biomass is assumed to result in negligible $CO_2$ emissions. The emission factors for power generation from fossil fuels are given by 771 gCO$_2$/kWh (oil), 353 gCO$_2$/kWh (CCGT), 488 gCO$_2$/kWh (OCGT), 1007 gCO$_2$/kWh (hard coal) and 1170 gCO$_2$/kWh (lignite). For a further description of the model and the underlying data see [16], [17].

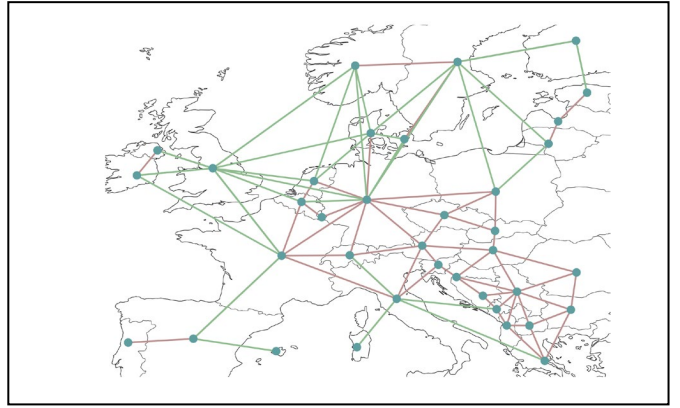

Figure 1. European power system model with 37 nodes. Red lines indicate AC connections, green lines indicate DC connections

The *base scenario* only optimizes the operation of the system given the existing generation and storage capacities as of 2021. Table 1 displays the total annual demand and the generation per type for this scenario, as well as the already installed generation capacity per type. It should be noted that due to the limited coverage of the underlying power plant list, the model only includes a part of the actual generation fleet in the real system (see the discussion in [16], [18], [19]). The average emission intensity of power generation in the system is given by 314.31 gCO$_2$/kWh. This value is higher than, for instance, the value of 255 gCO$_2$/kWh for the EU CO$_2$ intensity in 2022 as published in [20]. This difference in the absolute value results from limited coverage of the actual electricity system in the model, changes in the dispatch, or different emission factors for the different generation technologies. Here, we use the emission intensity for the base scenario as the basis for our evaluation of emission reductions due to either an emission cap or minimum renewable shares.

Table 1: CHARACTERISTICS OF THE BASE SCENARIO (LEFT) AND THE SCENARIO RESULTING FROM UNCONSTRAINED CAPACITY EXPANSION (RIGHT)

|  | Base scenario | | Unconstrained optimization | |
|---|---|---|---|---|
| Demand | 3215 TWh | | 3215 TWh | |
| Emission intensity | 314.31 gCO$_2$/kWh | | 285.04 gCO$_2$/kWh | |
|  | Capacity [GW] | Generation [TWh] | Capacity [GW] | Generation [TWh] |
| Hard coal | 63.9 | 422.8 | 63.9 | 381.5 |
| Lignite | 45.4 | 392.6 | 45.4 | 393.1 |
| Nuclear | 110.1 | 722.8 | 110.1 | 716.4 |
| CCGT | 184.3 | 347.0 | 184.3 | 208.8 |
| OCGT | 13.7 | 3.3 | 13.7 | 0.275 |
| Oil | 5.6 | 1.5 | 5.6 | 0.074 |
| Run of river | 49.4 | 168.9 | 49.4 | 169.1 |
| Biomass | 21.3 | 171.3 | 21.3 | 173.1 |
| Onshore wind | 184.6 | 368.1 | 201.4 | 425.1 |
| Offshore wind | 24.99 | 101.2 | 24.99 | 101.2 |
| Solar | 153.0 | 165.6 | 257.4 | 312.2 |
| Hydro storage | 155.4 | 337.4 | 155.4 | 349.0 |
| Battery storage | 0 | 0 | 0.04 | 0.050 |
| Hydrogen | 0 | 0 | 0.0008 | 0.0004 |

## III. RESULTS

As an extension of the base scenario, we implement an unconstrained capacity expansion optimization for the system. In this scenario, the system is allowed to cost-efficiently expand renewable generation capacities from solar, onshore wind and offshore wind, and invest in power generation from fossil gas (open cycle gas turbines, OCGT, and combined cycle gas turbines, CCGT) and grid extensions of existing connections up to 150% of the original transmission capacity. As storage capacities, both battery and hydrogen storage can be installed. Hydro storage, run of river, and all conventional generation other than fossil gas cannot be expanded, but are included in the operational optimization. Table 1 shows that already without minimum renewable shares or emission constraints, it is cost-efficient to expand renewable generation capacities from onshore wind and from solar. The resulting additional renewable generation alters the operation of the system, leading to changes in the dispatch of conventional generation capacities and an overall decline of the emission intensity of power generation by 9.3% compared to the base scenario.

In the following scenarios we either implement an emission cap or minimum renewable shares as boundary conditions for the expansion optimization. Minimum renewable shares are defined with a temporal granularity $\tau$ corresponding to hours (H), days (D), weeks (W), months (M), or the entire year (Y) under consideration. The set $S$ of renewable generation types includes wind onshore, wind offshore and solar as expandable types, plus run of river, which cannot be expanded. Biomass and nuclear power generation are not included in the set of renewable generation types in the given model but count as generation with zero carbon emissions. For a given renewable share with temporal granularity $\tau$, the capacity expansion optimization results in a system with a corresponding emission reduction. Equivalently, for a given emission cap (i.e. emission reduction), we can calculate the resulting minimum renewable share for each temporal granularity $\tau$.

Figure 2 visualizes this relation between minimum renewable shares of different temporal granularity and emission reductions under cost-efficient capacity expansion. The base system with a CO$_2$ intensity of 314 gCO$_2$/kWh results in a renewable share of 26% over the course of the year, or a minimum renewable share of 6% for hourly granularity, i.e. in each hour at least 6% of generation is provided from renewable generation (with storage allowing to shift generation of this type over time). If we consider cost-efficient capacity expansion without further constraints, the CO$_2$ intensity of power generation is reduced by 9% to 285 gCO$_2$/kWh due to increasing generation from onshore wind and solar (see Table 1). The corresponding minimum renewable share is increased to 31% over the entire year, or 9% for hourly granularity.

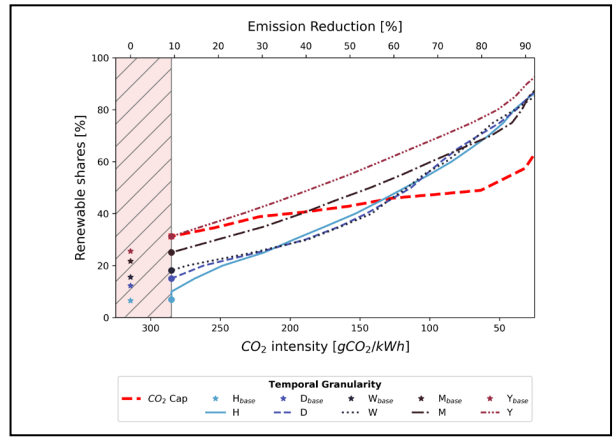

Figure 2. Relationship between reduced CO$_2$ intensity of power generation and corresponding minimum renewable shares with different temporal granularity. Stars represent the base scenario without capacity expansion. The red line displays the annual renewable share for given emission reductions. All other lines relate given minium renewable shares of different temporal granularity to the corresponding reduction in emissions.

The remaining part of Fig. 2 shows the necessary minimum renewable shares for each temporal granularity $\tau$ to achieve a given reduced CO$_2$ intensity of power generation. Emission reductions of up to 70% compared to the base scenario (CO$_2$ intensities of up to approximately 100 gCO$_2$/kWh) are achieved by similar minimum shares for temporal granularities of hours, days, and week. Given a granularity of months or the entire year, higher minimum shares are necessary to reach similar emission reductions, respectively. For even higher emission reductions close to 90% compared with the base scenario (CO$_2$ intensity of below 40gCO$_2$/kWh), very high minimum renewable shares for each temporal granularity are necessary. For a given emission cap, we also display the corresponding renewable share with yearly granularity. Figure 2 shows that starting with a share of approximately 31% for the unconstrained optimization scenario, the resulting yearly renewable share only increases to just below 50% for an emission reduction of 85% ($CO_2$ intensity of 50 $gCO_2$/kWh). At this point, the annual renewable shares as boundary conditions have to increase to almost 80% to yield this low overall $CO_2$ intensity. In our analysis, we exclude systems with close to zero $CO_2$ intensity, since in this region the nonconsideration of sector coupling leads to extreme results.

Figure 3 displays the generation mix and cost for capacity expansion for scenarios with a given emission cap, or for minimum renewable shares with yearly, weekly, or hourly temporal granularity. For scenarios with an emission cap (first row of Fig. 3), the emission reduction is mainly achieved due to a decrease in conventional generation (lignite, hard coal, oil), which is replaced by increasing generation from onshore wind and an increase in power generation from fossil gas. Only for high emission reductions (85%, or $CO_2$ intensity below 50 $gCO_2$/kWh) we observe the phaseout of fossil gas, and a corresponding uptake in renewable power generation. Capacity expansion costs (right column of Fig. 3) are dominated by investments in onshore wind, solar, and transmission capacity

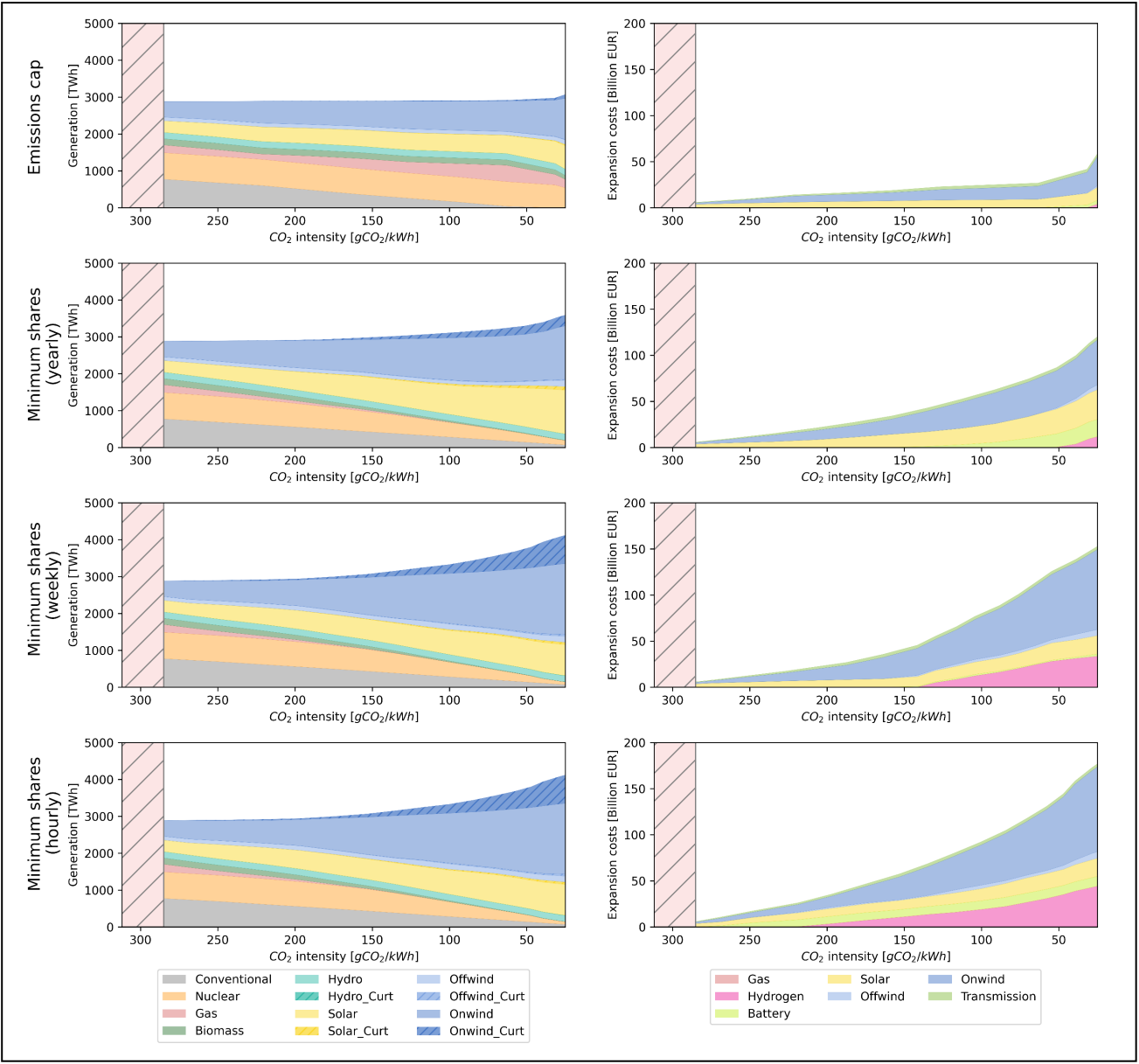

Figure 3. System characteristics for different scenarios. Left column shows generation mix, right column shows investment costs for capacity expansion. First row displays results for the system under an emission cap. Second to fourth row show results for scenarios with minimum renewable shares (yearly, weekly, hourly, respectively) as boundary conditions, resulting in an emission intensity as given on the x-axis.

Results for scenarios with minimum renewable are shown in the second to fourth row of Fig. 3. The left column displays the generation mix of a scenario with a certain emission reduction, resulting from a corresponding minimum renewable share with yearly (second row), weekly (third row) or hourly (fourth row) granularity.

For minimum renewable shares with yearly granularity, conventional generation is replaced by renewable generation from onshore wind and solar, without a fuel switch to fossil gas. Different from scenarios with an emission cap, nuclear power generation is also replaced by renewable generation, which is forced into the system by the boundary conditions of minimum renewable shares. The increase of renewable power generation incurs higher capacity costs, as displayed in the right column of Fig. 3. For yearly granularity, an emission reduction to a $CO_2$ intensity of 100 $gCO_2$/kWh, for instance, the necessary investment in generation capacity to fulfill the corresponding minimum renewable share on an annual basis is roughly twice as high as for the direct implementation through an emission cap.

For minimum renewable shares with an hourly granularity (fourth row of Fig. 3), the generation mix looks similar to scenarios based on shares with an annual granularity (note that the necessary minimum shares are higher for annual shares, as indicated in Fig. 2). The increase in renewable power generation in particular from onshore wind is higher, with some of the generation curtailed already for emission reductions by 50% compared to the base scenario. The capacity expansion costs for these scenarios with minimum hourly renewable shares are displayed in the right column of Fig. 3. Here, investment in battery and hydrogen storage already occurs at emission reductions of less than 50% compared to the base scenarios. These storage capacities are necessary to allow a temporal shift of renewable generation from wind and solar, which is necessary to fulfill the boundary condition of a certain minimum share of renewable power generation in each hour of the year. These additional investments lead to an increase in capacity costs compared to scenarios with minimum annual shares or an emission cap. For minimum shares with a weekly temporal granularity, the results (third row of Fig. 3) correspond to the temporal scale between yearly and hourly resolution.

IV. CONCLUSION AND OUTLOOK

In this contribution we studied the relationship between minimum renewable shares of different temporal granularity and the resulting emission reductions in an electricity system optimization model on the European scale. We also compared the resulting system design with scenarios where the emission reduction is directly implemented through an emission cap. Minimum renewable share with an hourly, daily, and weekly granularity lead to comparable emission reductions, whereas minimum shares with a monthly or annual granularity need significantly higher shares to achieve the same $CO_2$ intensity of power generation. For higher emission reductions (more than 80% compared to a base scenario), very high minimum shares are necessary for all temporal granularities. The system costs to achieve a given emission reduction target are considerably higher for minimum renewable shares compared to scenarios with an emission cap. By definition, in an optimization model an emission cap corresponds to the least cost solution to achieve a given emission reduction. This reduction is accomplished by investments in renewable power generation, but also due to a fuel switch to fossil gas and a continuing usage of low carbon power generation from nuclear. Minimum renewable shares, in contrast, enforce higher investments into renewable power generation to meet the given boundary conditions, which leads to higher cost for the resulting system. If the temporal granularity, for instance of an hour, additionally demands to shift renewable generation over time, the minimum renewable shares also lead to significant investments into storage capacities (battery and hydrogen storage).

These observations indicate that the system costs to achieve given emission reductions in scenarios with minimum renewable shares (corresponding to mechanisms like a temporally resolved feed-in tariff) are considerably higher than in the least-cost solution resulting from an emission cap (corresponding to an overall $CO_2$ price). Despite this drawback from the perspective of system costs, minimum renewable shares clearly incentivize higher investments into power generation from wind and solar and thus could be interpreted as more robust than the least cost solution, which, for instance, also makes use of a fuel switch to fossil gas. Minimum renewable shares with a higher temporal granularity additionally incentivize investments into flexibility options like battery and hydrogen storage.

The findings presented in this contribution only provide a starting point for further research. In reality, there will always be a mix of different mechanisms, so the combination of minimum shares and different emission prices should be taken into consideration. Also the spatial perspective needs to be included, which could be realized by additional constraints based on the geographic or grid distance between generation and consumption.

ACKNOWLEDGMENT

Tim Fürmann acknowledges funding from DFG (SPP 1984), project ID 450860949.